\documentclass[sigconf, screen]{acmart}

\setcopyright{acmcopyright}
\copyrightyear{2021}
\acmYear{2021}
\acmDOI{}

\usepackage[acronym]{glossaries}
\glsdisablehyper
\usepackage{makecell}
\usepackage{array}
\usepackage{graphicx}
\usepackage{colortbl}
\usepackage{multirow}
\usepackage{ragged2e}
\usepackage{hhline}
\usepackage{rotating}
\usepackage{bigstrut}

\newacronym{VR}{VR}{virtual reality}
\newacronym{MR}{MR}{mixed reality}
\newacronym{AR}{AR}{augmented reality}
\newacronym{XR}{XR}{extended reality}
\newacronym{UX}{UX}{user experience}
\newacronym{PX}{PX}{player experience}
\newacronym{UI}{UI}{user interface}
\newacronym{AI}{AI}{artificial intelligence}
\newacronym{EEG}{EEG}{electroencephalography}
\newacronym{DOF}{DoF}{degrees of freedom}
\newacronym{DDA}{DDA}{dynamic difficulty adjustment}
\newacronym{CAD}{CAD}{computer"~aided design}
\newacronym{IMU}{IMU}{inertial measurement unit}
\newacronym{WIM}{WIM}{World In Miniature}
\newacronym{SUS}{SUS}{System Usability Scale}
\newacronym{SUSPQ}{SUS PQ}{Slater\hbox{-}Usoh\hbox{-}Steed Presence Questionnaire}
\newacronym{PQ}{PQ}{presence questionnaire}
\newacronym{SSQ}{SSQ}{Simulator Sickness Questionnaire}
\newacronym{IPQ}{IPQ}{Igroup Presence Questionnaire}
\newacronym{IEQ}{IEQ}{Immersive Experience Questionnaire}
\newacronym{PXI}{PXI}{Player Experience Inventory}
\newacronym{E2I}{E2I}{Engagement, Enjoyment, and Immersion}
\newacronym{GengQ}{GengQ}{Game Engagement Questionnaire}
\newacronym{PDA}{PDA}{personal digital assistant}
\newacronym{EMR}{EMR}{electromagnetic resonance}
\newacronym{HCI}{HCI}{human-computer interaction}
\newacronym{SD}{SD}{standard deviation}
\newacronym{SLR}{SLR}{systematic literature review}
\newacronym[plural=RQs,firstplural=research questions (RQs)]{RQ}{RQ}{research question}
\newacronym[plural=CAVEs,firstplural=cave automatic virtual environments (CAVEs)]{CAVE}{CAVE}{cave automatic virtual environment}
\newacronym[plural=RPGs,firstplural=role\hbox{-}playing games (RPGs)]{RPG}{RPG}{role\hbox{-}playing game}
\newacronym{STEM}{STEM}{Science, Technology, Engineering and Mathematics}
\newacronym[plural=HMDs,firstplural=head"~mounted displays (HMDs)]{HMD}{HMD}{head\hbox{-}mounted display}
\newacronym[plural=RCPs,firstplural=rich client platforms (RCP)]{RCP}{RCP}{rich client platform}
\newacronym[plural=IVs,firstplural=independent variables (IV)]{IV}{IV}{independent variable}
\newacronym[plural=DVs,firstplural=dependent variables (DV)]{DV}{DV}{dependent variable}

\begin{document}

\title{Questionnaires and Qualitative Feedback Methods to Measure User Experience in Mixed Reality}

\author{Tobias Drey}
\email{tobias.drey@uni-ulm.de}
\affiliation{%
  \institution{Institute of Media Informatics\\ Ulm University}
  \city{Ulm}
  \country{Germany}
}

\author{Michael Rietzler}
\email{michael.rietzler@uni-ulm.de}
\affiliation{%
  \institution{Institute of Media Informatics\\ Ulm University}
  \city{Ulm}
  \country{Germany}
}

\author{Enrico Rukzio}
\email{enrico.rukzio@uni-ulm.de}
\affiliation{%
  \institution{Institute of Media Informatics\\ Ulm University}
  \city{Ulm}
  \country{Germany}
}

\renewcommand{\shortauthors}{Drey et al.}

\begin{abstract}
Evaluating the user experience of a software system is an essential final step of every research.
Several concepts such as flow, affective state, presences, or immersion exist to measure user experience.
Typical measurement techniques analyze physiological data, gameplay data, and questionnaires.
Qualitative feedback methods are another approach to collect detailed user insights.
In this position paper, we will discuss how we used questionnaires and qualitative feedback methods in previous mixed reality work to measure user experience.
We will present several measurement examples, discuss their current limitations, and provide guideline propositions to support comparable mixed reality user experience research in the future.
\end{abstract}




\maketitle

\section{Introduction}
Evaluating the \gls{UX} is vital in \gls{HCI} research, especially when developing new artifact contributions (see Wobbrock and Kientz~\cite{10.1145/2907069}) such as new interaction systems~\cite{drey2020vrsketchin} or games~\cite{10.1145/3334480.3382789}.
Several concepts were introduced during the last years to measure user states.
Flow introduced by Csikszentmihalyi~\cite{csikszentmihalyi1990flow,csikszentmihalyi1997finding,csikszentmihalyi1975beyond} describes a state of optimal experience with eight conditions necessary to reach it.
Affective state with its components emotion, mood, and core affect~\cite{ekkekakis2012affect} is another example.
Especially emotions such as anxiety~\cite{chen_self-regulation_2016}, boredom~\cite{liu_effect_2011}, enjoyment~\cite{10.1145/2556288.2557078}, or frustration~\cite{10.1145/3242671.3242682} were used in previous works to measure \gls{UX}\hbox{-}related states.
According to Schrader et al.~\cite{Schrader2017}, measures for these emotions can be grouped into the categories \textit{physiological data}, \textit{gameplay data}, and \textit{questionnaires}.
Presence, the feeling of being there~\cite{heeter1992being}, and immersion, a vivid illusion of the reality~\cite{doi:10.1162/pres.1997.6.6.603}, are further methods to describe \gls{UX}.
Questionnaires for them exist as well (e.g., \gls{PQ} by Witmer and Singer~\cite{doi:10.1162/105474698565686} or \gls{E2I} questionnaire~\cite{996519}) and were previously used to evaluate systems~\cite{10.1145/3173574.3173702,10.1145/3332165.3347871}.
Many of these concepts overlap (see Mekler et al.~\cite{10.1145/2556288.2557078}), and \gls{MR} systems add further challenges~\cite{citeWorkshop}.

Besides physiological data, gameplay data, and questionnaire measures, qualitative feedback methods such as interviews or usability walkthroughs exist as well~\cite{drey2020vrsketchin,10.1145/3196709.3196755}.
They can provide detailed insights but need a valid method to reduce subjectivity~\cite{doi:10.1080/14780887.2020.1769238}.

This work will give advice on how \gls{UX} can be measured by sharing the authors' experiences from their previous \gls{MR} works~\cite{drey2020vrsketchin,10.1145/3173574.3174034,8613757,10.1145/3196709.3196755,10.1145/3173574.3173702,10.1145/3139131.3139145,10.1145/3332165.3347871,10.1145/3313831.3376821}.
It will focus on \textit{questionnaires} and \textit{qualitative \gls{UX} measures} and will discuss the limitations of both approaches to gather individual user feedback.
Furthermore, we will provide guideline propositions to support comparable \gls{MR} \gls{UX} research in the future.

\section{Questionnaires and Qualitative Feedback Methods}
This section shows our personal experience regarding questionnaires and qualitative feedback methods for \gls{UX} evaluation and discusses the used methods.
\subsection{Questionnaires}
To evaluated the \gls{UX} in our previous works~\cite{drey2020vrsketchin,10.1145/3196709.3196755,10.1145/3173574.3173702,10.1145/3139131.3139145,10.1145/3332165.3347871,10.1145/3313831.3376821}, we used questionnaires for presence (\gls{SUSPQ}~\cite{doi:10.1162/105474600566989}, \gls{PQ} by Witmer and Singer~\cite{doi:10.1162/105474698565686}), immersion and enjoyment (\gls{E2I}~\cite{996519}), simulator sickness (\gls{SSQ}~\cite{996519}), and general usability (\gls{SUS}~\cite{brooke1996sus}).
Further questionnaires were introduced during the past years, such as the \gls{IPQ}\footnote{http://www.igroup.org/pq/ipq/index.php}, the \gls{IEQ}~\cite{JENNETT2008641}, or the \gls{PXI}~\cite{ABEELE2020102370}, to name just a few.
They all use generalizable questions and are limited to their specific \gls{UX} state.
None of these were explicitly designed for the evaluation of \gls{MR} systems.
To measure the \gls{MR} artifact specific\hbox{-}aspects, we therefore often created custom single\hbox{-}item questions to address them~\cite{10.1145/3173574.3174034,8613757,10.1145/3196709.3196755,10.1145/3332165.3347871,10.1145/3313831.3376821}.

This non\hbox{-}exhaustive overview shows that many questionnaires exist to measure \gls{UX}\hbox{-}related aspects, and for the same experience state (for presence e.g., \gls{IPQ}, \gls{SUSPQ}, \gls{PQ} by Witmer and Singer).
Nevertheless, current research lacks a generalizable questionnaire to measure the overall \gls{MR} \gls{UX}.
Therefore, researchers always need to choose the appropriate questionnaires for their study and create additional custom single\hbox{-}item questions.
This variety of standardized and custom questionnaires makes it difficult and often impossible to compare different publications.

Another issue of these questionnaires is that they often have multiple subscales, which (1) do not always apply to every situation or (2) overlap with other questionnaires.
An example for (1) is the \gls{PQ} by Witmer and Singer, which has a subscale regarding auditory cues that is not always applicable.
For example, this is the case for our work VRSketchIn~\cite{drey2020vrsketchin}, which had no audio feedback.
This presence questionnaire would therefore be not applicable, or only with some subscales, to evaluate VRSketchIn.
An example for (2) is the \gls{PXI} which has subscales regarding immersion (e.g., overlapping with the \gls{IEQ}) and ease of control (e.g., overlapping with the \gls{SUS}).

An important question is also how the questionnaires should be presented correctly, retrospectively after the experience or during it.
In most of the works, we decided to put the questionnaires outside the experience. 
However, in retrospect, integrating questionnaires into the application offers many advantages.
For example, there is no break in the experience, and participants can answer the questions in the world they are supposed to evaluate~\cite{citeWorkshop}.

Therefore, we propose that guidelines should be created that describe when and how these questionnaires should be used to make their results more reliable and more comparable between publications and propose in this work guideline propositions for discussion at the workshop.

\subsection{Qualitative Feedback Methods}
In addition to questionnaires, we used other qualitative feedback methods regarding \gls{UX} in previous works~\cite{drey2020vrsketchin,10.1145/3196709.3196755}.
To evaluate the \gls{UX} of VRSketchIn~\cite{drey2020vrsketchin}, we used a usability walkthrough and encouraged our participants to state feedback and ask questions.
A similar approach was used by us for evaluating VRSpinning~\cite{10.1145/3196709.3196755}.

These approaches were very well suited to get detailed insight into how our systems were used by the participants and helped us improve them.
As both were \gls{VR} systems, we had to simultaneously be able to record the user itself in the physical world and the view of the user inside \gls{VR}.
Both pieces of information were necessary to understand their behavior and interactions, analyze their feedback, and draw the right conclusions.

Qualitative feedback methods such as usability walkthroughs have obstacles regarding their scientific evaluation and tend to be subjective.
To reduce subjectivity, approaches like Grounded Theory~\cite{strauss1994grounded,glaser2017discovery} or Thematic Analysis~\cite{doi:10.1191/1478088706qp063oa,doi:10.1080/14780887.2020.1769238} exist.
These approaches are time\hbox{-}consuming and error\hbox{-}prone if not applied correctly~\cite{doi:10.1080/14780887.2020.1769238,https://doi.org/10.1046/j.1365-2648.2000.01430.x}.
It is furthermore complicated to objectively compare two independent systems with usability walkthroughs as the researchers always interpret the users' statements.

Therefore, we propose that guidelines should be defined how qualitative evaluation methods should be used for artifact (according to Wobbrock and Kientz~\cite{10.1145/2907069}) evaluation.
These guidelines\linebreak should further consider questionnaires and other \gls{UX} measures previously shown in the introduction and provide specific instructions for \gls{MR} research.
We will provide guideline propositions in the following for discussion at the workshop.

\section{Guideline Propositions for Measuring User Experience}
To measure and report \gls{UX} in \gls{MR} systems and based on our previously stated experiences, we propose the following guidelines:
\begin{description}
\item[1.] \textit{When using questionnaires with subscales that were not designed for \gls{MR} systems, only the applicable subscales should be considered.}\newline
Questionnaires such as the \gls{PQ} have subscales that are not always applicable, e.g., for audio or haptics, which not every system has, and in this case, should not be used.
\item[2.] \textit{When using multiple questionnaires, overlapping questions and subscales should be discussed.}\newline
When the \gls{PXI} and the \gls{IEQ} questionnaire are used together, both measure immersion (\gls{PXI} with a subscale).
The used questions, however, are different and, therefore, measure immersion differently.
This may cause different results, which should be discussed.
\item[3.] \textit{When using measures such as presence or immersion, the used definition should be referenced.}\newline
For measures such as presence or immersion, different definitions and questionnaires exist (presence, definitions:~\cite{heeter1992being,doi:10.1162/105474698565686,sheridan1992musings,miall2002marie}, questionnaires: \gls{IPQ}, \gls{SUSPQ}, \gls{PQ} by Witmer and Singer; immersion, definitions:~\cite{doi:10.1162/105474698565686,doi:10.1162/pres.1997.6.6.603,mcmahan2003immersion}, questionnaires: \gls{IEQ}, \gls{E2I}, \gls{PXI}).
To allow comparison with other works, the used definition and questionnaires should always be referenced.
\item[4.] \textit{When using questionnaires with subscales, their results should be reported.}\newline
To make a paper as comparable to other works as possible, the results of subscales of questionnaires should always be reported as sometimes other works report only subscales of a questionnaire. 
\item[5.] \textit{When using single\hbox{-}item questions, their exact wording should be reported.}\newline
To make these custom questions as transparent and understandable as possible, their exact wording should be reported.
\item[6.] \textit{When artifacts or systems are evaluated exploratory, a qualitative feedback method can enhance quantitative findings.}\newline
Even when using multiple questionnaires and other quantitative measures, it is difficult to get a sophisticated as well as a holistic impression of the \gls{UX}.
Qualitative feedback methods such as think\hbox{-}aloud usability walkthroughs or\linebreak semi\hbox{-}structured interviews can provide this.
To reduce subjectivity, a standardized evaluation method should be used for them.
\item[7.] \textit{When artifacts or systems are evaluated exploratory, the users' actions should be recorded for retrospective comparison with quantitative and qualitative results.}\newline
Recordings provide an additional level of understanding during an evaluation and help to set the participants' answers and results into context.
\end{description}

\section{Conclusion}
We have shown that different methods exist to measure \gls{UX} and described how we used questionnaires and qualitative feedback methods in previous work.
Both have limitations, and further research has to be conducted to provide guidelines on when and how to use which \gls{UX} measurement method, especially for \gls{MR} research.
We think that such guidelines will increase the overall evaluation results of publications and allow a more reliable comparison of evaluation results between them.
To start a discussion, we have created guideline propositions.

At the workshop, the presenting author wants to (1) exchange with the other participants their experiences using \gls{UX} measures, (2) discuss how the evaluation workflow in \gls{MR} studies can be improved for these measures, and (3) discuss our guideline propositions for \gls{MR} \gls{UX} evaluation methods to allow inter\hbox{-}system and inter\hbox{-}research comparisons.

\begin{acks}
This work has been conducted within the DFG project ”Empirical Assessment of Presence and Immersion in Augmented and Virtual Realities“ (RU 1605/4-1).
\end{acks}

\bibliographystyle{ACM-Reference-Format}
\bibliography{sample-authordraft}

\end{document}